\documentclass[intlimits,twoside,a4paper]{article}

\usepackage{amsmath,amssymb}
\usepackage{graphicx}

\usepackage[T2A]{fontenc}
\usepackage[cp1251]{inputenc}

\usepackage[eqsecnum]{cmpj2}

\issue{2013}{16}{4}{43603}
\doinumber{10.5488/CMP.16.43603}

\title[Interaction of the model alkyltrimethylammonium ions with alkali halide salts]%
{Interaction of the model alkyltrimethylammonium ions with alkali halide salts:
an explicit water molecular dynamics study\thanks{It is a pleasure to dedicate this paper to our good friend and
coworker Professor M.F.~Holovko.}}
\author[M.~Druchok, \v C. Podlipnik, V.~Vlachy]{M.~Druchok\refaddr{label1},
        \v C. Podlipnik\refaddr{label2},
        V.~Vlachy\refaddr{label2}}
\addresses{
\addr{label1} Institute for Condensed Matter Physics, 1 Svientsitskii St., 79011 Lviv, Ukraine
\addr{label2} Faculty of Chemistry and Chemical Technology, University of Ljubljana, 5 A\v sker\v ceva St., 1000 Ljubljana, Slovenia
}

\authorcopyright{M.~Druchok, \v C. Podlipnik, V.~Vlachy, 2013}
\date{Received July 24, 2013, in final form August 29, 2013}

\begin{document}

\maketitle

\begin{abstract}
We present an explicit water molecular dynamics simulation of
dilute solutions of model alkyltrimethylammonium surfactant
ions (number of methylene groups in the tail is 3, 5, 8, 10,
and 12) in mixture with NaF, NaCl, NaBr, and NaI salts,
respectively. The SPC/E model is used to describe water
molecules. Results of the simulation at 298~K are presented in
the form of radial distribution functions between nitrogen and
carbon atoms of CH$_2$ groups on the alkyltrimethylammonium
ion, and the counterion species in the solution. The
running coordination numbers between carbon atoms of
surfactants and counterions are also calculated. We show that I$^-$
counterion exhibits the highest, and F$^-$ the lowest affinity
to ``bind'' to the model surfactants. The results are discussed
in view of the available experimental and simulation data for
this and similar solutions.
\keywords surfactants, alkyltrimethylammonium salts, alkali halides, ion binding, molecular dynamics
\pacs 61.20.Ja, 61.20.Qg, 82.20.Wt, 82.30.Rs, 82.35.Rs
\end{abstract}

\section{Introduction}

It has long been known that properties of surfactant solutions,
for example critical micelle concentration, depend on the
nature of the counterion present in
solution~\cite{Birch,Delisi1988,Rozycka2001,Perger,Jakubowska2008,Sarac,Jakubowska2010}.
The latter effect being of particular interest to us, it is further
examined herein using the explicit water molecular dynamics method.
In view of our present study, important experimental results
were presented by Ro\v zycka-Roszak et al.~\cite{Rozycka2001}.
This group performed calorimetric measurements (see figure~1 of
their paper) showing that upon dissociation of the
dodecyltrimethylammonium bromide micelle, the heat is consumed.
In contrast to this, the dissociation of
dodecyltrimethylammonium chloride micelles is exothermic, i.e., heat
is released. The experiments also indicated that the two halide
ions influence the process of micelle formation in a completely
different way.

The situation closely resembles the one observed in recent
experimental studies of aliphatic
$x,y$-ionenes~\cite{Cebasek2011,Boncina2012,Serucnik2012}.
Ionenes are cationic polyelectrolytes with a different ($x,y$
may vary from 3,3 to 12,12) number of methylene groups between
the quaternary ammonium groups. Notice that aliphatic ionenes
are similar in chemical structure to the alkyltrimethylammonium
ions studied here. For example, the monomer unit of the
12,12-ionene is actually the dodecyltrimethylammonium ion. The
heats of dilution of the $x,y$-ionenes indicate a strong
difference between the salts with different counterions. The
results for $x,y$-ionene fluorides are exothermic and
rather well follow  the theoretical results based on the continuum
solvent models. The salts with bromide, chloride, and iodide
anions as counterions show different
behavior~\cite{Cebasek2011,Arh}: heat of dilution may be either
endothermic or exothermic depending on the length of the
hydrophobic portion of the charge. Moreover, the most recent
experimental data indicated~\cite{Cebasek2013} that additions
of extra methylene groups (transition from 6,12- to
12,12-ionene) affect the solutions with different counterions
differently.

Among the theoretical methods of probing the hydration in the
explicit water models, the molecular dynamics simulations seem to be the most useful (see, for example,
references~\cite{Lounnas,mf1,mf2,Virtanen}). In the present explicit
water molecular dynamics study we are interested in the effects of
an increasing number of methylene groups (i.e., the length of the
surfactant tail) on the interaction between the various
counterion species in a solution and quaternary ammonium group of
the surfactant. For this purpose, we performed the study of
mixtures of alkyltrymethylammonium surfactant ions with
fluoride, chloride, bromide, and iodide counterions in mixture
with NaF, NaCl, NaBr, and NaI salts, respectively. The
simulation is focused on very dilute solutions with respect to
surfactant. In this way, the counterion-charged group
interaction can be studied without possible complications of the
polyion chain association. The study accordingly applies to the
conditions well below the critical micelle concentration. The
surfactant molecules with the tail, containing from three to
twelve CH$_2$ groups, are probed.

\section{Model and simulation details}

\begin{figure}[!b]
\begin{center}
\includegraphics[width=0.8\textwidth]{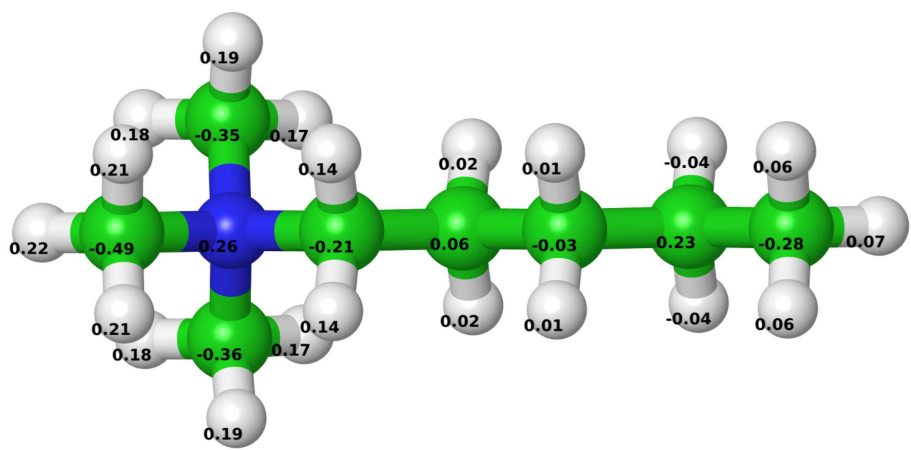}
\caption{(Color online) Schematic representation of the alkylammonium ion with five methyl groups in the tail.
Nitrogen is indicated by blue, carbons by green, and hydrogens by white color. The numbers show optimized atomic charges.}
\label{fig:aa}
\end{center}
\end{figure}

A set of surfactant solutions containing alkylammonium ions,
sodium co-ions, neutralizing halide counterions, and water
molecules are investigated. To obtain insights into the specific
ion effects, we examined a series of halide salts: fluorides,
chlorides, bromides, and iodides. The respective counterions
are characterized by the same charge but different crystal size,
so one can expect a somewhat different interaction of the
counterion with the surfactant (quaternary ammonium) ion. The
latter ion consists of the quaternary nitrogen, neighbored by
three CH$_3$-groups, a carbon chain with CH$_2$ groups and a
terminal CH$_3$ group. As an example, in figure~\ref{fig:aa} we
show the surfactant ion with five carbons in the tail.
In the present study we considered surfactants with five
different tail lengths; i.e., with 3-, 5-, 8-, 10-, and 12
carbons denoted as C$_3$, C$_5$, C$_8$, C$_{10}$, and
C$_{12}$. The wide spectrum of chain lengths allows us to study
the effect of hydrophobicity on the counterion-quaternary
ammonium interaction.

All the particles in the system were treated on equal footing.
Water was described within the SPC/E model~\cite{SPC}. The
Lennard-Jones parameters for Na$^+$ ions were taken from
reference~\cite{aqvist}, the ones for the halide ions~--- from the
work by Palinkas~\cite{palinkas}. The Lennard-Jones parameters
for the sites of a surfactant were taken from the OPLS force
field~\cite{Jorgensen1996}. Charges at the atoms of surfactants
were fitted from the electrostatic potential calculated with DFT
B3LYP/6-311G** using Jaguar~7.9 (Schr\"odinger
Suite)~\cite{blyp}. Listing these charges in the form of tables
for the whole atom simulation is far from optimal. As an
example, figure~\ref{fig:aa} shows the distribution of charges for
C$_5$, while all interested
readers are encouraged to request the full information from
authors via e-mail. In order to preserve the intramolecular
geometry of the surfactants, we have utilized a set of bond and
angular potentials in the form $U=k_\textrm{bond}(r-r_0)^2$ and
$U=k_\textrm{angle}(\alpha-\alpha_0)^2$. The carbon-carbon and
nitrogen-carbon distances $r_0$ in the surfactant backbone are
1.5~{\AA} with the corresponding
$k_\textrm{bond}=500$~kcal/mol/{\AA}$^2$, the parameters for the
backbone angles~--- $\alpha_0=111^\circ$,
$k_\textrm{angle}=250$~kcal/mol. The parameters for the
carbon-hydrogen bonds in CH$_2$ and CH$_3$ groups are
250~kcal/mol/{\AA}$^2$ and 1.1~{\AA}. For the
hydrogen-hydrogen distances inside CH$_2$ and CH$_3$ groups we
used additional constraints with parameters
150~kcal/mol/{\AA}$^2$, 1.8~{\AA}. As for two off-backbone CH$_3$
groups, the corresponding carbon-nitrogen-carbon angle is
controlled by the angular potential
$250(\alpha-109^\circ)$~kcal/mol. Finally, for the surfactant
rigidity we also applied a set of angular potentials
$250(\alpha-180^\circ)$~kcal/mol between next but one backbone
carbons (first-third-fifth atom, second-fourth-sixth, \ldots).

The Lennard-Jones parameters ($\sigma_i$, $\epsilon_i$) assigned
to various atoms or ions are shown in table~\ref{Tab1}.
\begin{table}
\caption{Model parameters.}
\label{Tab1}
\begin{center}
\begin{tabular}{|c|c|c|c|}\hline
 & species & $\epsilon$ (kcal/mol) & $\sigma$ ({\AA}) \\
\hline\hline
water & O & 0.1554 & 3.1656 \\
      & H & 0.0 & 0.0 \\
\hline
       & N & 0.170 & 3.25 \\
AA     & C & 0.066 & 3.50 \\
       & H & 0.030 & 2.5 \\
\hline
            & Na & 0.0028 & 3.3304 \\
            & F &  0.0118 & 4.0 \\
electrolyte & Cl & 0.0403 & 4.86 \\
            & Br & 0.0645 & 5.04 \\
            & I &  0.0979 & 5.40 \\
\hline
\end{tabular}
\end{center}
\end{table}
For unlike sites, the
parameters were obtained using the mixing rules in
the form $\sigma_{ij} = \frac{1}{2}(\sigma_i+\sigma_j$) and
$\epsilon_{ij} = \sqrt{\epsilon_i \epsilon_j}$.
A standard DL\_POLY~\cite{dlpoly} package was used for molecular
dynamics simulations of model solutions with the unit cell
containing 2352 water molecules, one surfactant ion, one
counterion X (X can be F$^-$,Cl$^-$, Br$^-$, or I$^-$), and 6
Na$+$X pairs modelling the added salt. The concentration $c_s$ of the
added low-molecular electrolyte Na$^+$X$^-$ was $\approx
0.14$~mol dm$^{-3}$. All the species were allowed to move
freely across the cubic cell with periodic boundary conditions.
The long-range Coulomb interactions were taken into account by
the Ewald summation technique. The short-range interactions
ware truncated at $R_\textrm{cut}=15$~{\AA}. The pressure (1~bar) and
temperature (298~K) were controlled by means of a Nose-Hoover
barostat and thermostat in an isotropic $N$, $P$, $T$ ensemble within
the Melchiona modification~\cite{Melchionna}. The number of
steps in the production runs ranged from 3.5 to 4.0$\times
10^7$ with the time step $5 \times 10^{-16}$~s. Such a time
step is needed to satisfy the constraints imposed by the bond
and angular intramolecular potentials for water and
surfactants. The radial distribution functions~(RDF) between
various sites on the molecule and ions in solution are
presented in the form of graphs.

\section{Numerical results}

\subsection{Radial distribution functions}

\begin{figure}[!t]
\centerline{
\includegraphics[clip=true,width=0.45\textwidth]{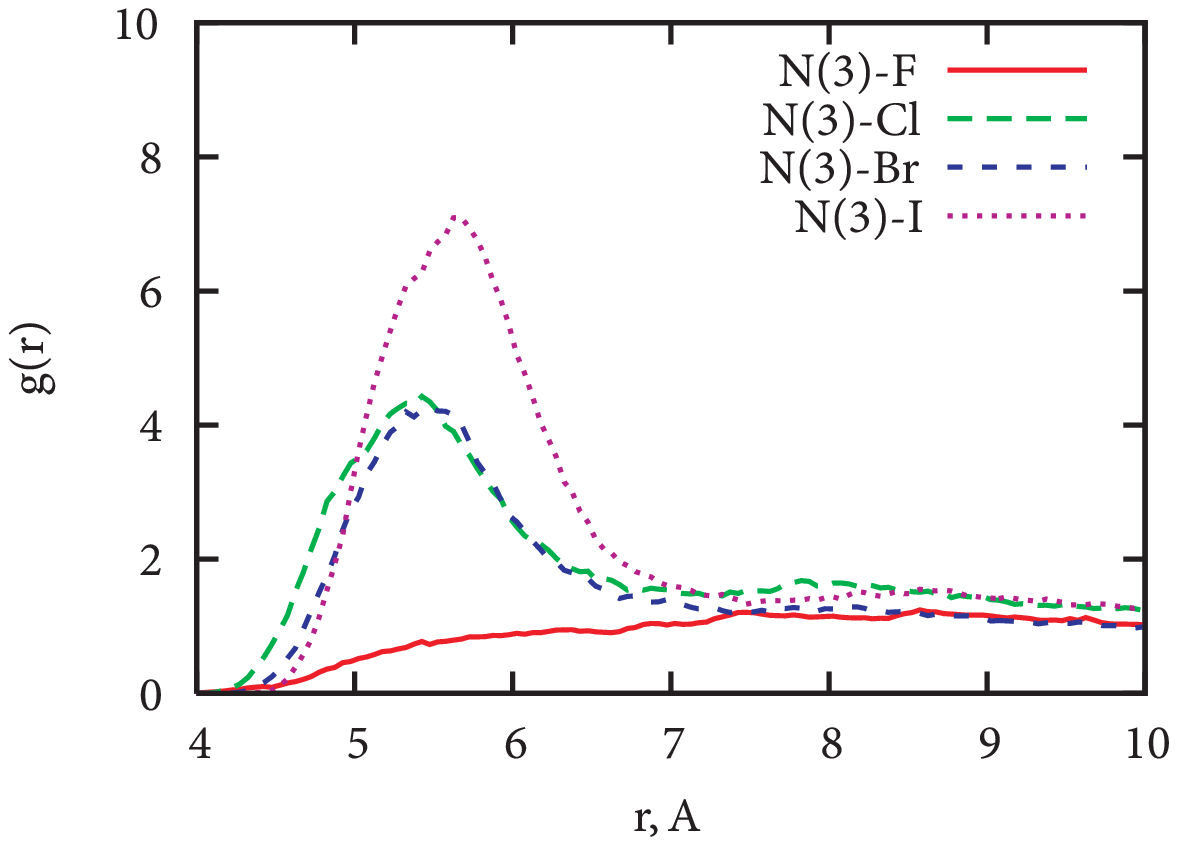}
\includegraphics[clip=true,width=0.45\textwidth]{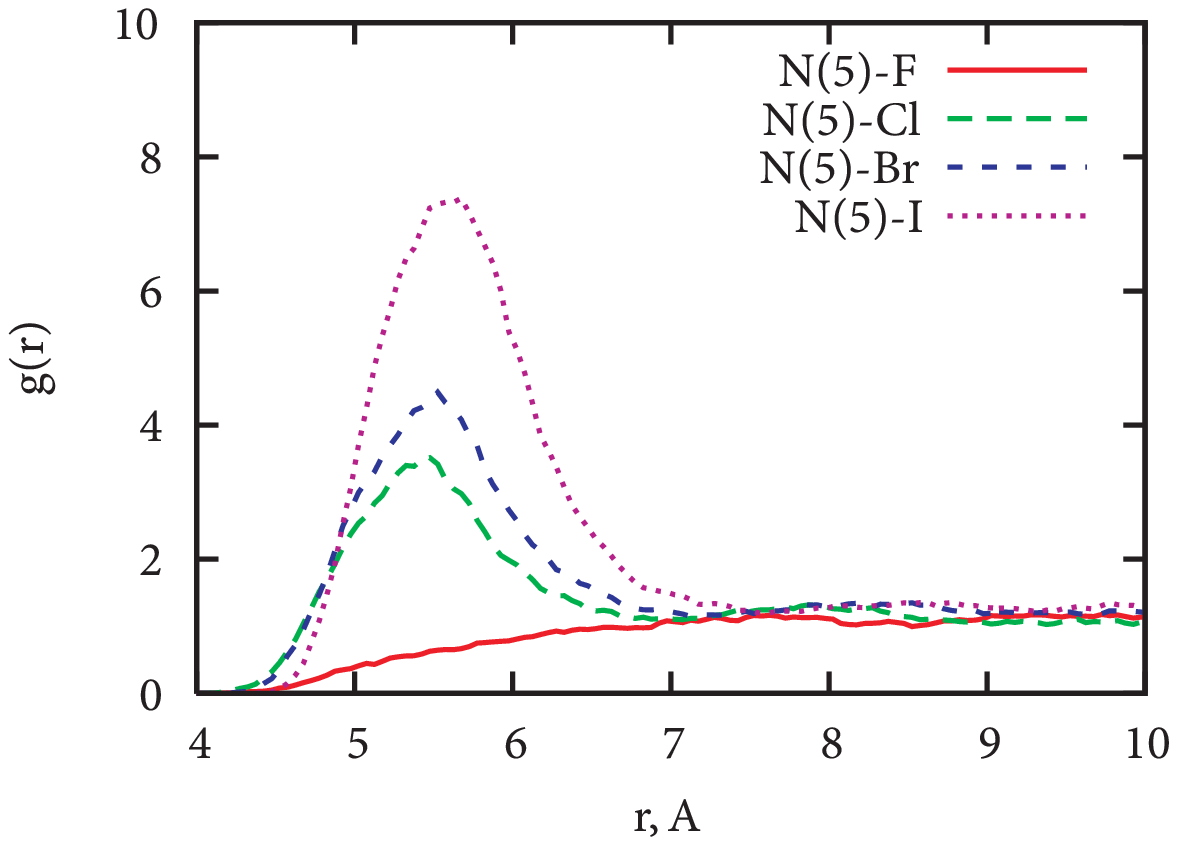}}
\centerline{
\includegraphics[clip=true,width=0.45\textwidth]{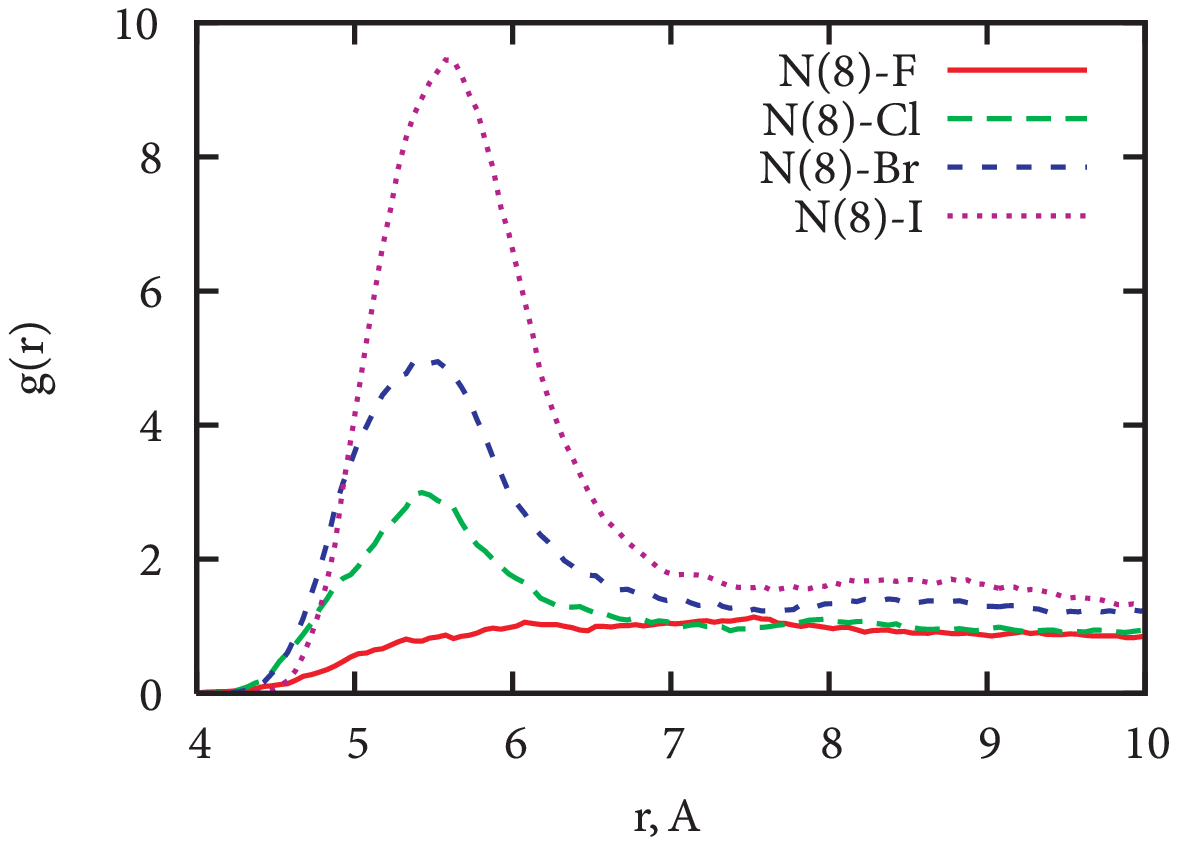}
\includegraphics[clip=true,width=0.45\textwidth]{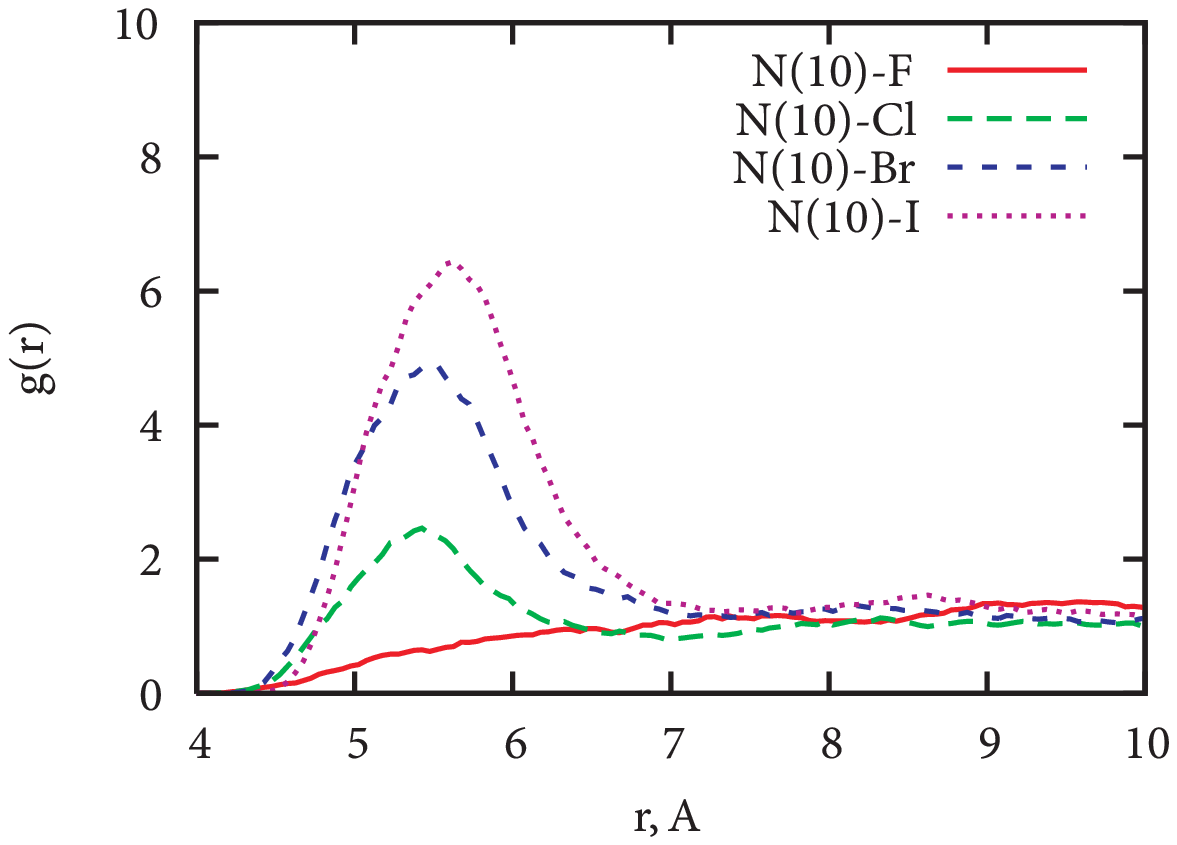}
}
\centerline{
\includegraphics[clip=true,width=0.45\textwidth]{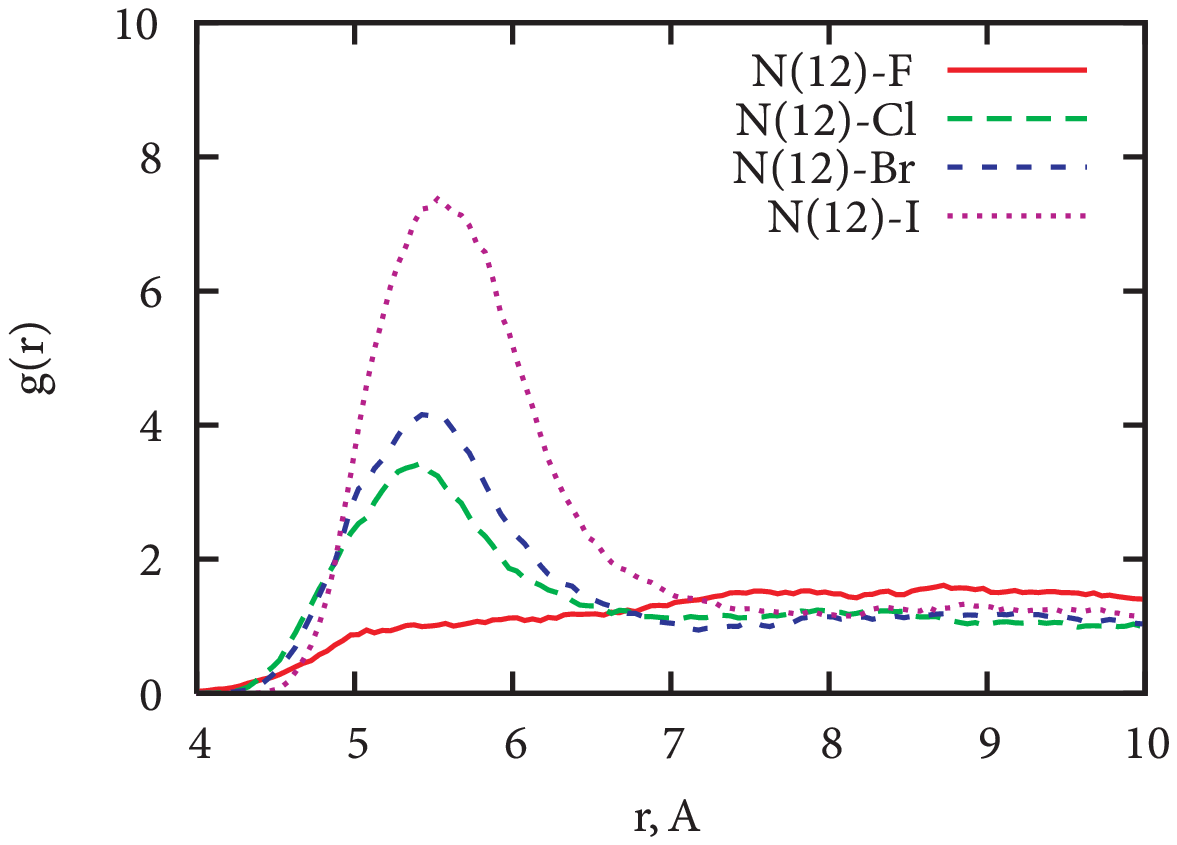}
}
\caption{(Color online) Nitrogen-counterion RDFs for molecule with (a) three methylene groups, C$_3$,
(b) C$_5$, (c) C$_8$, (d) C$_{10}$, and (e) C$_{12}$. The results for
fluorides are denoted by full red lines, for chlorides by dashed
green lines, bromides are denoted by short-dashed blue, and
iodides by dotted magenta lines.}
\label{fig:n-x}
\end{figure}

First, in figure~\ref{fig:n-x} we consider the
nitrogen-counterion radial distribution functions. Notice that
in all the solutions studied here an extra sodium salt (NaX,
X$^-$ is the counterion) is present.
The plots show that fluoride counterions tend to be located in
the bulk of the solution (away from the surfactant ion),
demonstrating poor correlation with the quaternary ammonium
ion. This behavior is a consequence of strong hydration of
fluoride ion, preventing ``association'' of fluoride ions with
the nitrogen group on the surfactant, as we already know from the
experimental~\cite{Cebasek2011,Boncina2012,Serucnik2012} and
theoretical~\cite{Druchok2008,Druchok2009} studies of
$x,y$-ionene solutions.
\begin{figure}[!t]
\centerline{
\includegraphics[clip=true,width=0.45\textwidth]{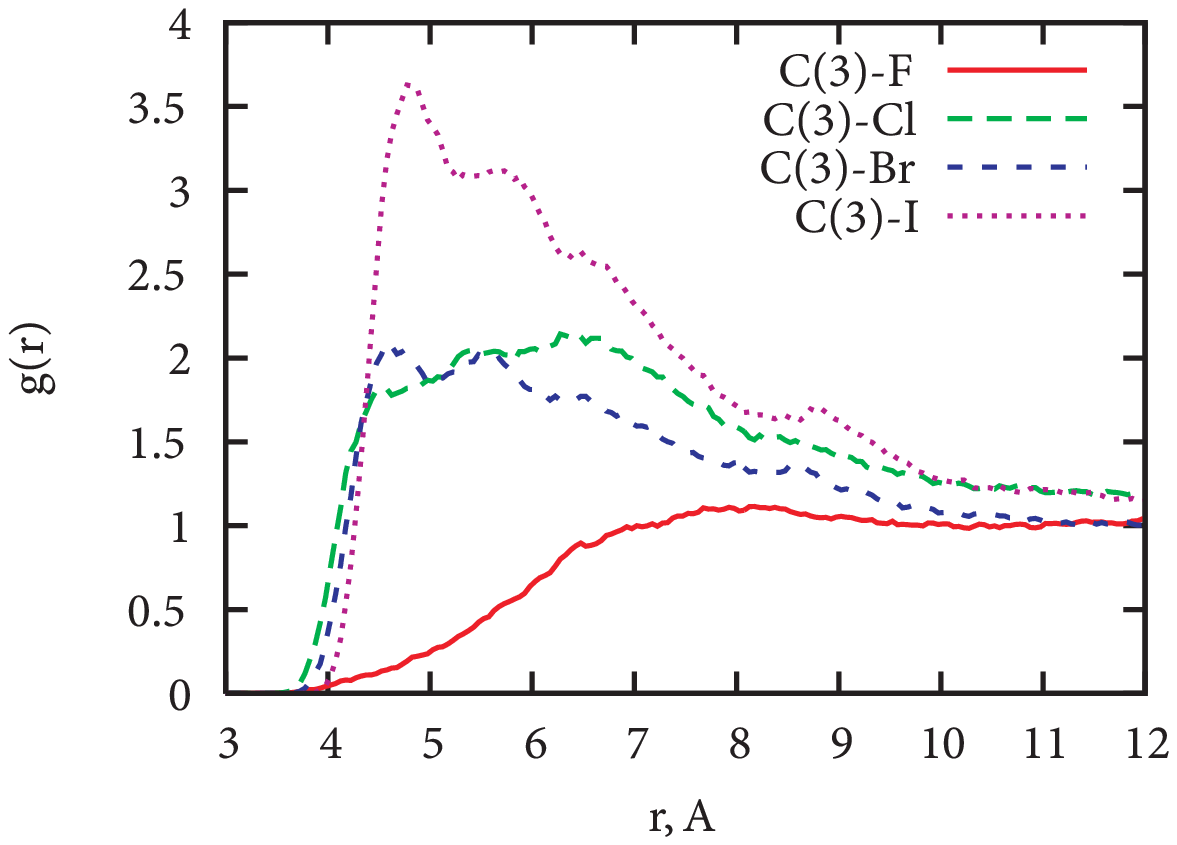}
\includegraphics[clip=true,width=0.45\textwidth]{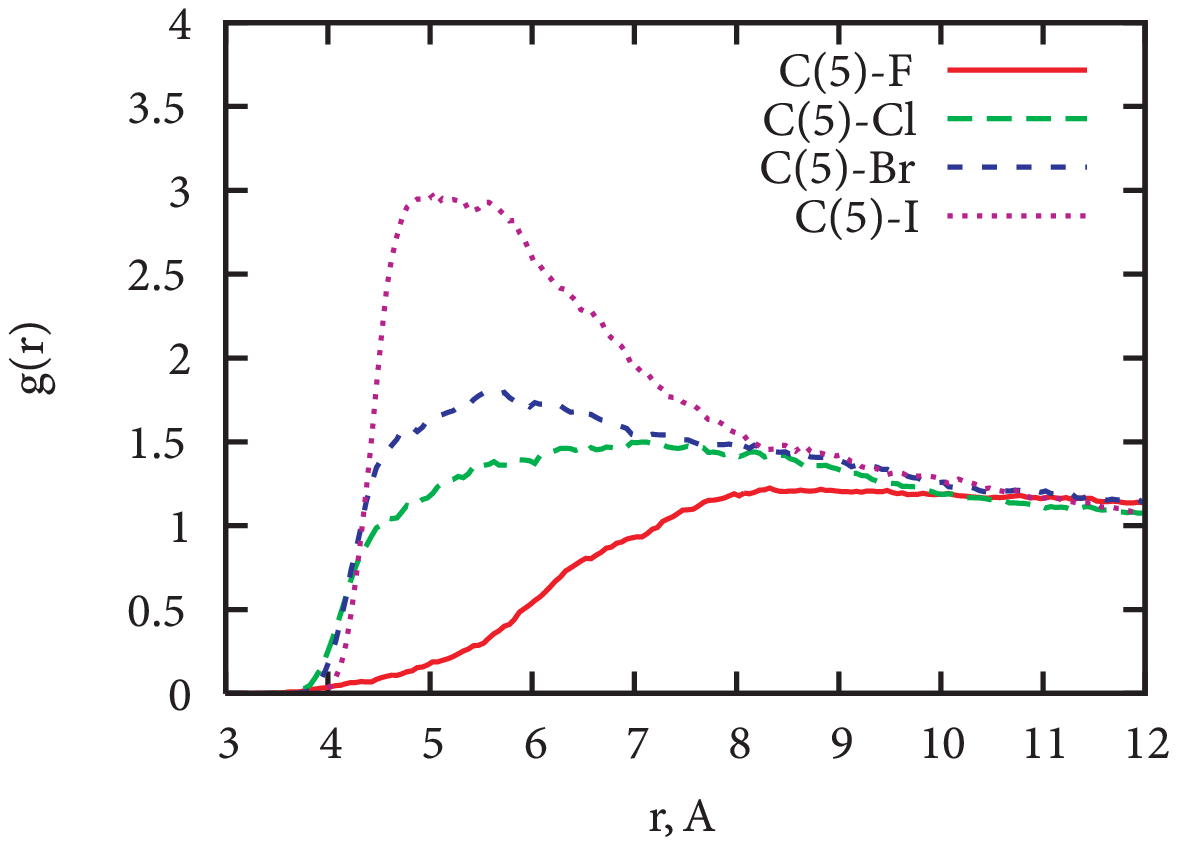}
}
\centerline{
\includegraphics[clip=true,width=0.45\textwidth]{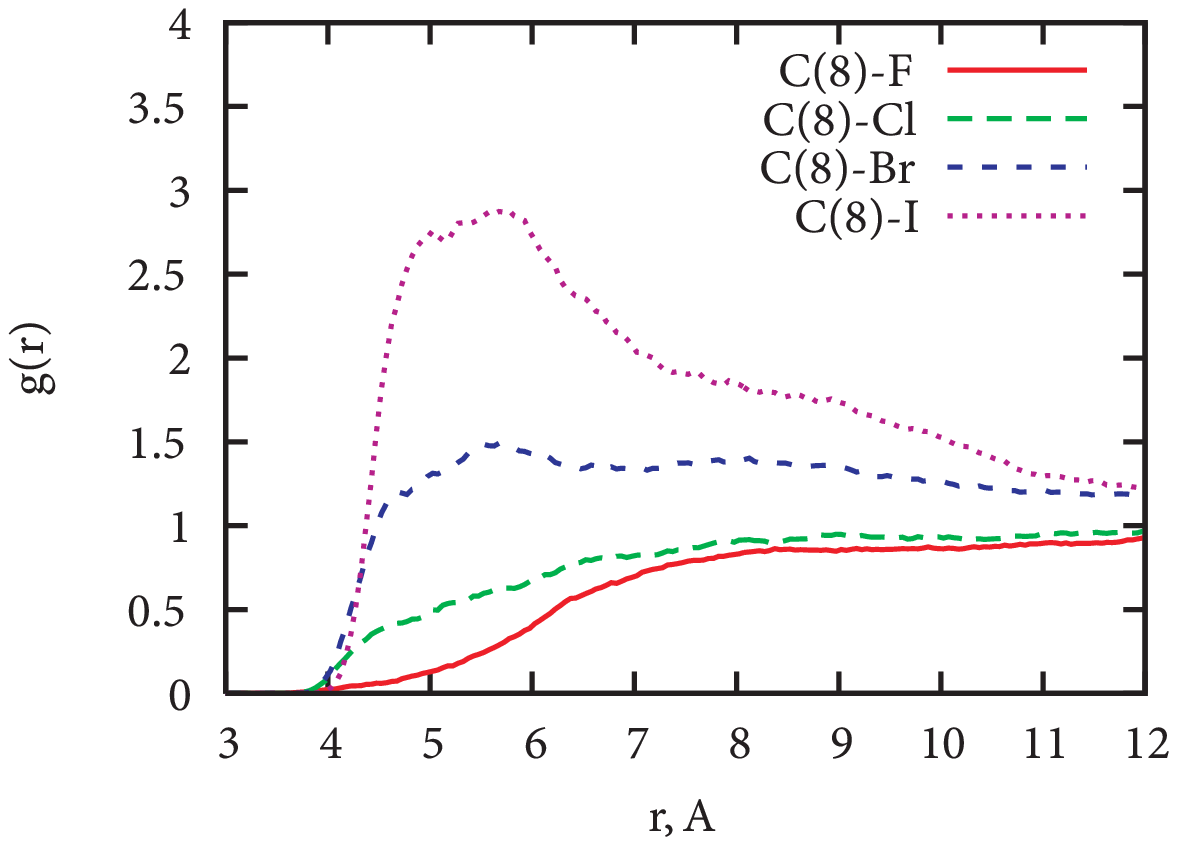}
\includegraphics[clip=true,width=0.45\textwidth]{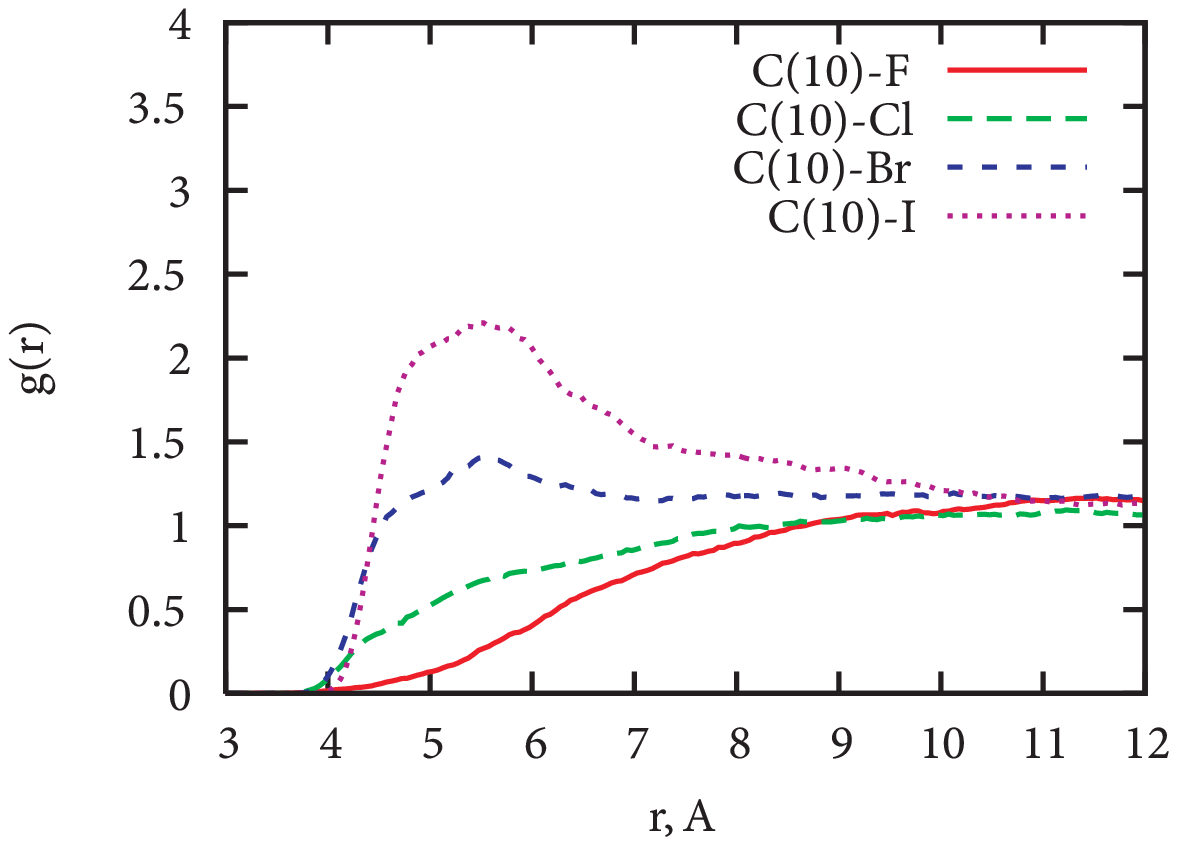}
}
\centerline{
\includegraphics[clip=true,width=0.45\textwidth]{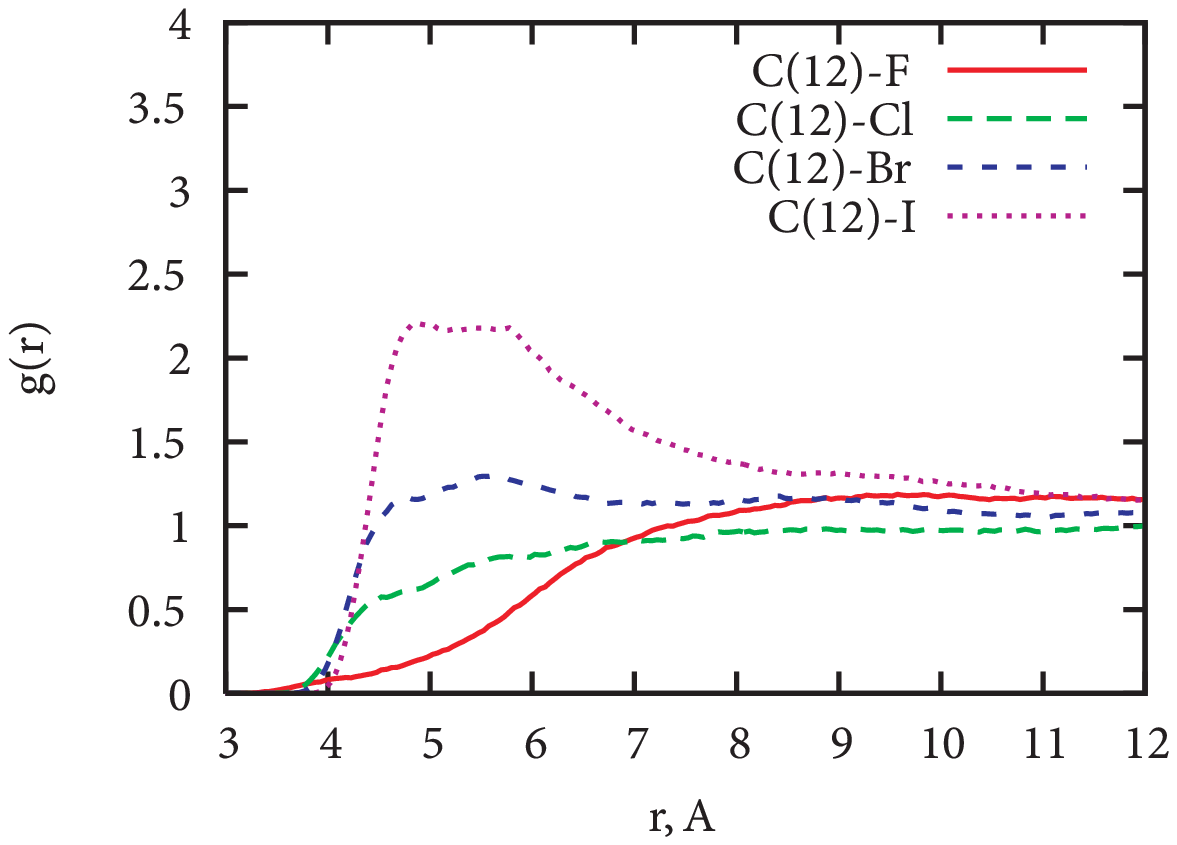}
}
\caption{(Color online) Carbon-counterion radial distribution functions.  The legend as
for figure~\ref{fig:n-x}.}
\label{fig:c-x}
\end{figure}
The halides of larger sizes, such as chloride, bromide, and
iodide counterions, are less strongly hydrated (their free
energies of hydration are smaller in magnitude than that of
fluoride ion) and can, therefore, release some water molecules in
interaction with the quaternary nitrogen group on the
surfactant. This causes the nitrogen-counterion peak height to
follow the ordering: I$^-$ $>$ Br$^-$ $>$ Cl$^-$
$>$ F$^-$.
This is valid for all the surfactants studied here, an
exception being the shortest one, C$_{3}$, where the peak heights
of the chloride and bromide ions are approximately the same.

Different halide ions follow different sequences of the first
peak height with respect to the chain length: for strongly
solvated chlorides,  the shortest, C$_{3}$, surfactant has the
highest peak. The peak height then decreases in the order
$\rm{C}_3 > \rm{C}_5 > \rm{C}_{12} > \rm{C}_8 >
\rm{C}_{10}$.
In contrast with chlorides, the iodide salts, which are known to
release their hydration waters more easily, reveal a different
dependence of the magnitude of the first RDF peak. The highest
value is obtained for C$_{8}$: $\rm{C}_8 > \rm{C}_5 \approx \rm{C}_{12}
> \rm{C}_3 > \rm{C}_{10}$. Here, the number of methylene
groups becomes more important; a higher number of CH$_2$ groups
means a stronger attraction. The balance between the Coulomb (charge
density decreases with the chain length) and van der Waals
attraction, which increases with the number of methylene groups
in the chain, produces the observed sequence. Notice that,
according to literature, the iodide ion is considered to have
more affinity for hydrophobic surfaces than other halide
ions~\cite{Lund2008}. The peak sequence for bromides reveals
the lowest value for C$_{12}$, the others run close with a slight
prevalence of the C$_8$ one, like it is found for the iodide
solutions.

These results are complemented by the carbon-counterion radial
distribution functions~(figure~\ref{fig:c-x}). Notice that only
the carbons from the surfactant tails are
included into this RDF calculation; the contribution of carbons of CH$_3$ groups
neighboring to nitrogens is neglected. The sequence of
carbon-counterion peak heights is: I$^-$ $>$ Br$^-$ $>$ Cl$^-$
$>$ F$^-$. Again an exception is the shortest molecule (C$_3$),
where the peak heights of the chloride and bromide counterions
are approximately the same. Peaks of these distribution
functions are generally lower than those of the
counterion-nitrogen distributions, indicating a weaker
accumulation of counterions next to the surfactant tail. Similar to the
nitrogen-fluoride radial distribution functions, the
carbon-fluoride ones exhibit merely a weak correlation,
confirming the conclusion about marginal ``association'' of
F$^-$ ions with the surfactant molecules. The carbon-chloride
distribution functions demonstrate a quite expected sequence of
the RDF peak heights: the highest is C$_3$ peak, next goes
C$_5$, then $\rm{C}_8 \approx \rm{C}_{10} \approx \rm{C}_{12}$.
This indicates that the relatively strongly solvated chloride
counterions exhibit little (or no) affinity for surfactants
with longer tails. The carbon-iodide pair distribution
functions demonstrate a higher peak for C$_3$ than
$\rm{C}_5 \approx \rm{C}_8$, while the lowest are the peaks for
$\rm{C}_{10}$ and $\rm{C}_{12}$. The counterion-carbon
interaction, as judged on the basis of these RDFs, is stronger
for I$^-$ than for Cl$^-$, and this effect is quite prominent for longer surfactants~\cite{Lund2008}.

\clearpage

\subsection{Running coordination numbers}

\begin{figure}[!b]
\centerline{
\includegraphics[clip=true,width=0.45\textwidth]{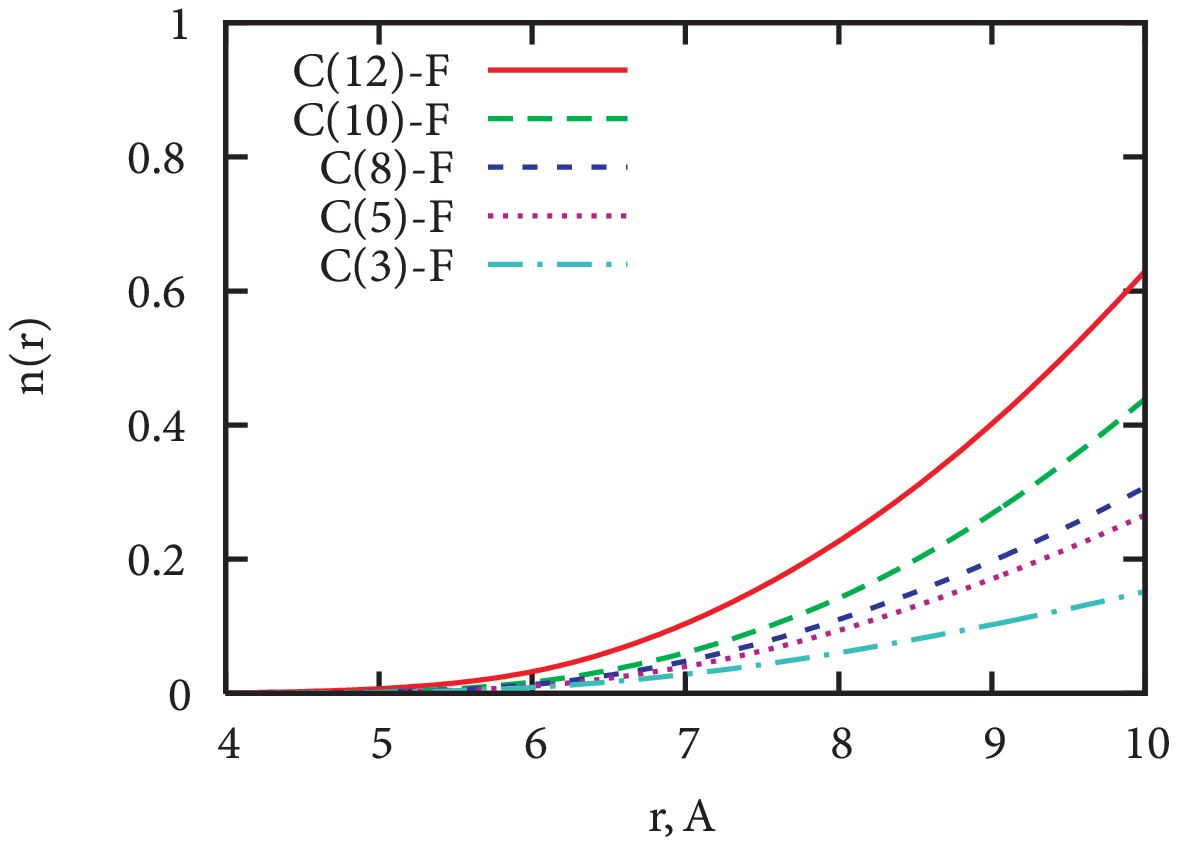}
\includegraphics[clip=true,width=0.45\textwidth]{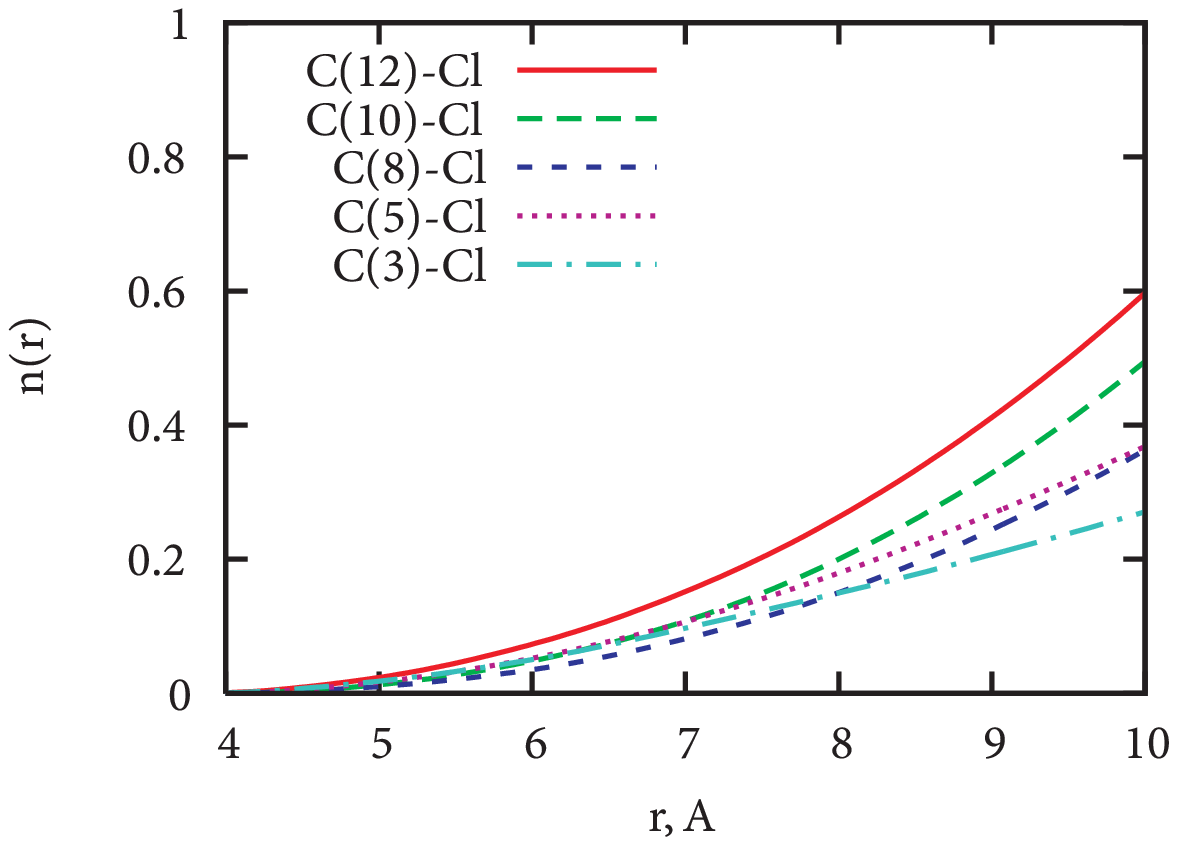}
}
\centerline{
\includegraphics[clip=true,width=0.45\textwidth]{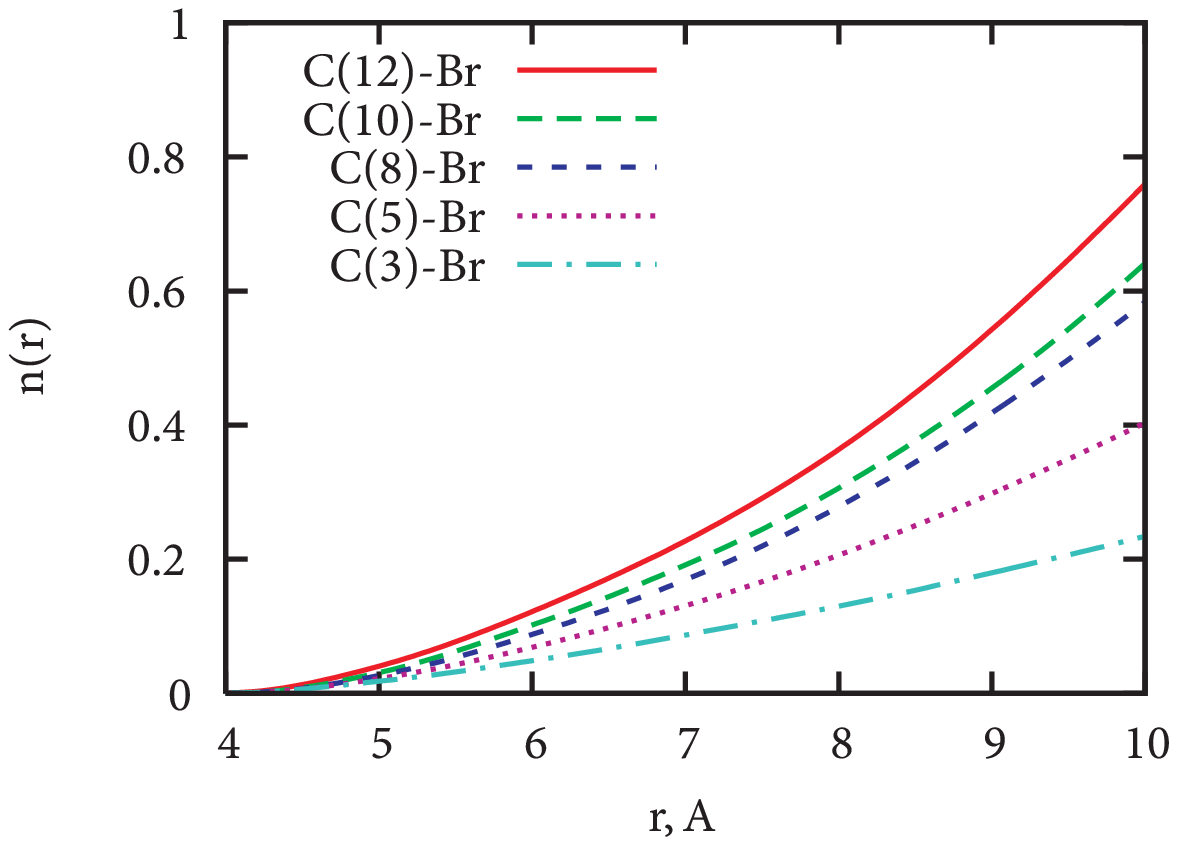}
\includegraphics[clip=true,width=0.45\textwidth]{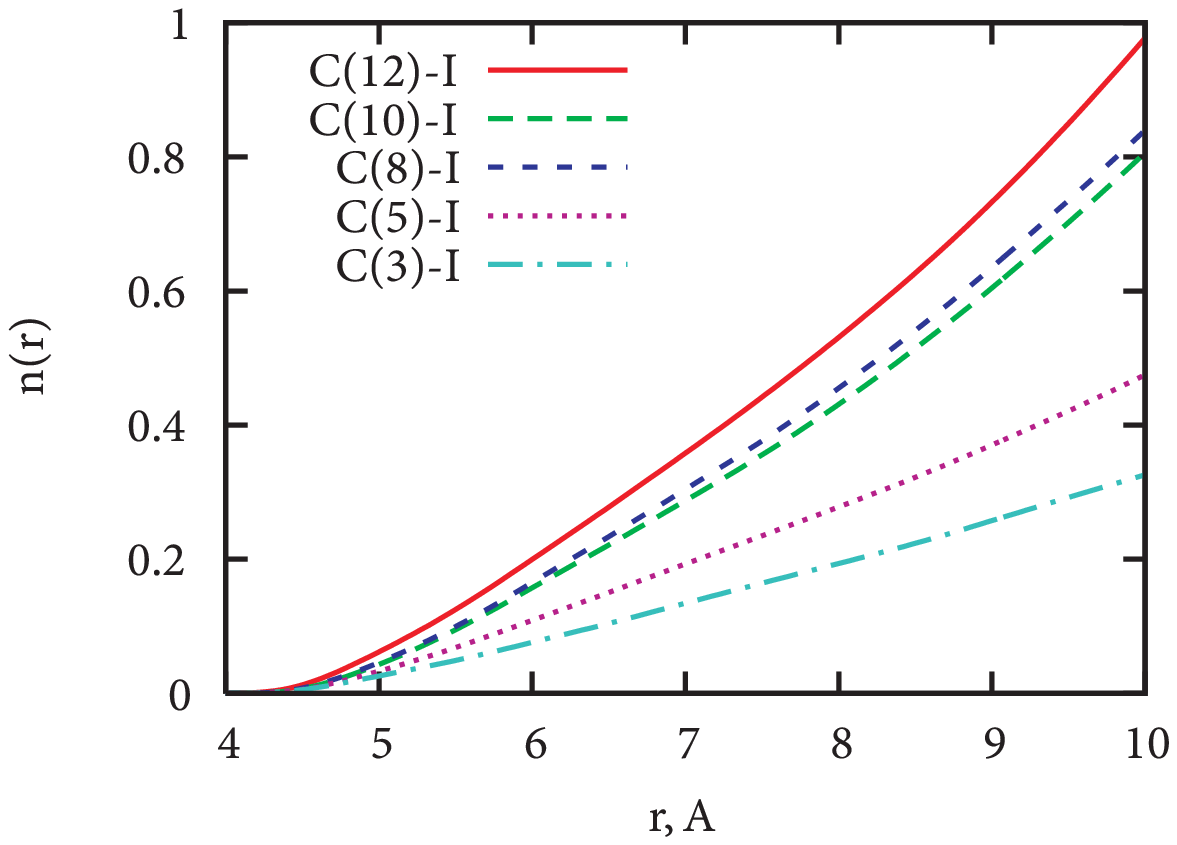}
}
\caption{(Color online) Counterion-carbon running coordination numbers for the
fluorides~(top left), chlorides~(right), bromides~(bottom left),
and iodides~(right hand panel). C$_{12}$ results are denoted by full red
lines, C$_{10}$~--- dashed green, C$_8$~--- short-dashed blue,
C$_5$~--- dotted magenta, and C$_3$ results are denoted by dash-dotted
cyan lines.}
\label{fig:c-x_n}
\end{figure}

Each of the surfactant species carries one charged group but
a different number of methylene groups (C$_x$), which
obscures the direct comparison (nitrogen-counterion vs
carbon-counterion) of the RDFs discussed above. For this
reason, we also present the so-called running coordination
numbers as an alternative way of data analysis. This
quantity, $n_{\alpha\beta}(r)$, is defined as a number of
particles $\beta$ in a sphere of radius $r$ around a certain
particle $\alpha$ in the center:
\begin{equation}
n_{\alpha\beta}(r)=4\pi\rho_{\beta}\int_0^r
g_{\alpha\beta}(r')r'^2 \rd r', \label{coord}
\end{equation}
where
$\rho_{\beta}$ is a number density of species $\beta$ in the
system. We denote the counterion species with subscript $\alpha$ and carbon
of the methylene group with $\beta$, so the coordination numbers
reflect the probability to find a carbon atom in the sphere
around a counterion (figure~\ref{fig:c-x_n}). This permits a
more realistic presentation of the counterion accumulation
around the methylene groups, as it can be inferred from the
relevant pair distribution functions alone.
The conclusion emerging from figure~\ref{fig:c-x_n} is that
fluoride ions are the least, and iodide counterions the most
strongly coordinated to carbon atoms, while the other
counterions fit in-between. Generally, the running coordination
numbers for C$_{12}$ are the highest, with a decrease in the
direction toward C$_3$. The present observation reflects the fact that
C$_{12}$ solutions contain more carbons than the C$_3$ ones.
The running coordination numbers for fluorides demonstrate a
monotonous dependence on the length of the hydrophobic chain.

For less hydrated chlorides, the picture complicates:
the C$_8$ running coordination number (for a region of
distances up to 8~{\AA}) runs bellow the other corresponding
functions. One should also note a relatively small difference
between C$_{12}$ and C$_3$ running coordination numbers for the
fluoride and chloride solutions. This finding is consistent
with the suggestion drawn on the basis of the
nitrogen-fluoride/chloride RDFs: the origin of the attraction
between surfactants and counterions is mainly of
electrostatic nature. The picture changes if one considers
bromide and iodide solutions. The gap between C$_{12}$ and
C$_3$ values becomes more significant, moreover, for the iodide
case, the C$_8$ running coordination numbers exceed C$_{10}$
values. This agrees with the jump seen above in the
C$_8$-iodide RDFs (figure~\ref{fig:n-x}). We ascribe this
behavior to the competition between Coulomb and van der Waals
interactions; while the importance of the former decreases with an
increasing chain length, just an opposite is true for the van
der Waals forces. We have also calculated the
counterion-nitrogen running coordination numbers. These data
seem to merely confirm the insights obtained from the radial
distribution functions presented in figure~\ref{fig:n-x} and are,
therefore, not shown here.

\subsection{Hydration of the quaternary ammonium and methylene groups}

\begin{figure}[!t]
\centerline{
\includegraphics[clip=true,width=0.45\textwidth]{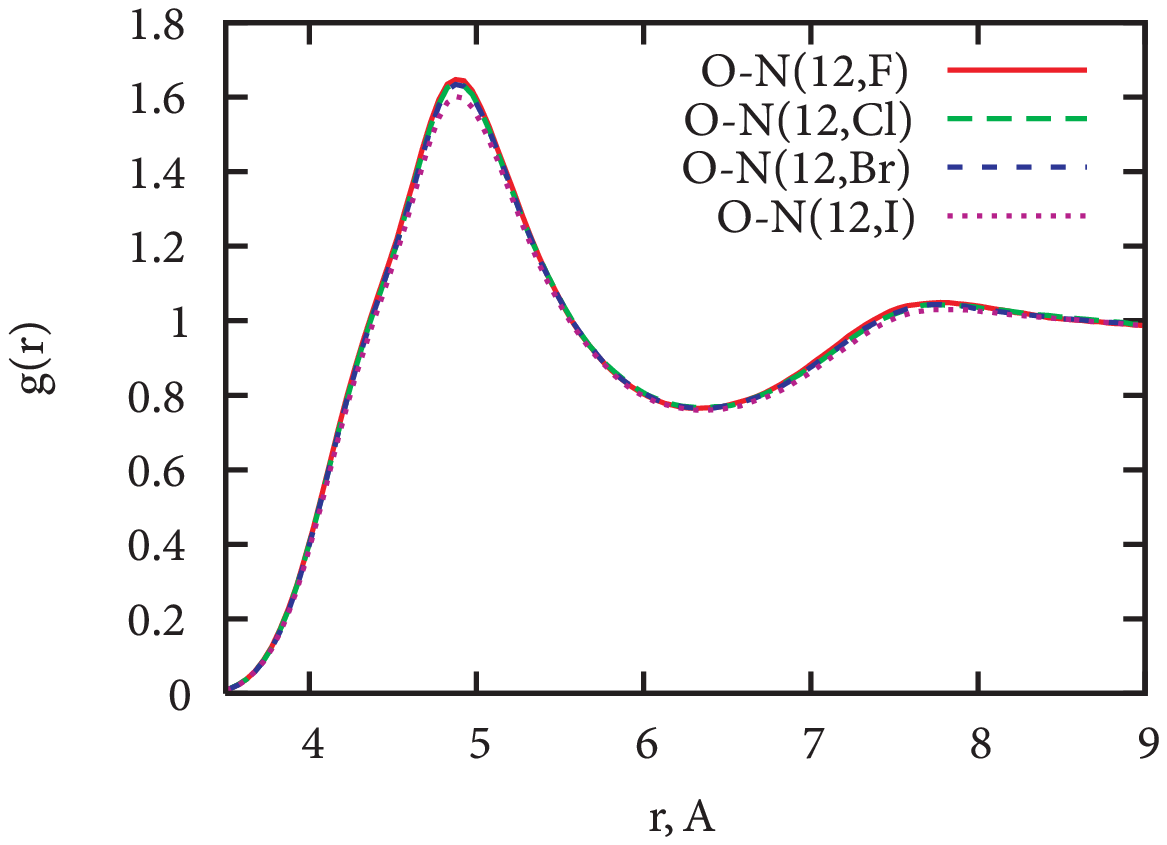}
\includegraphics[clip=true,width=0.45\textwidth]{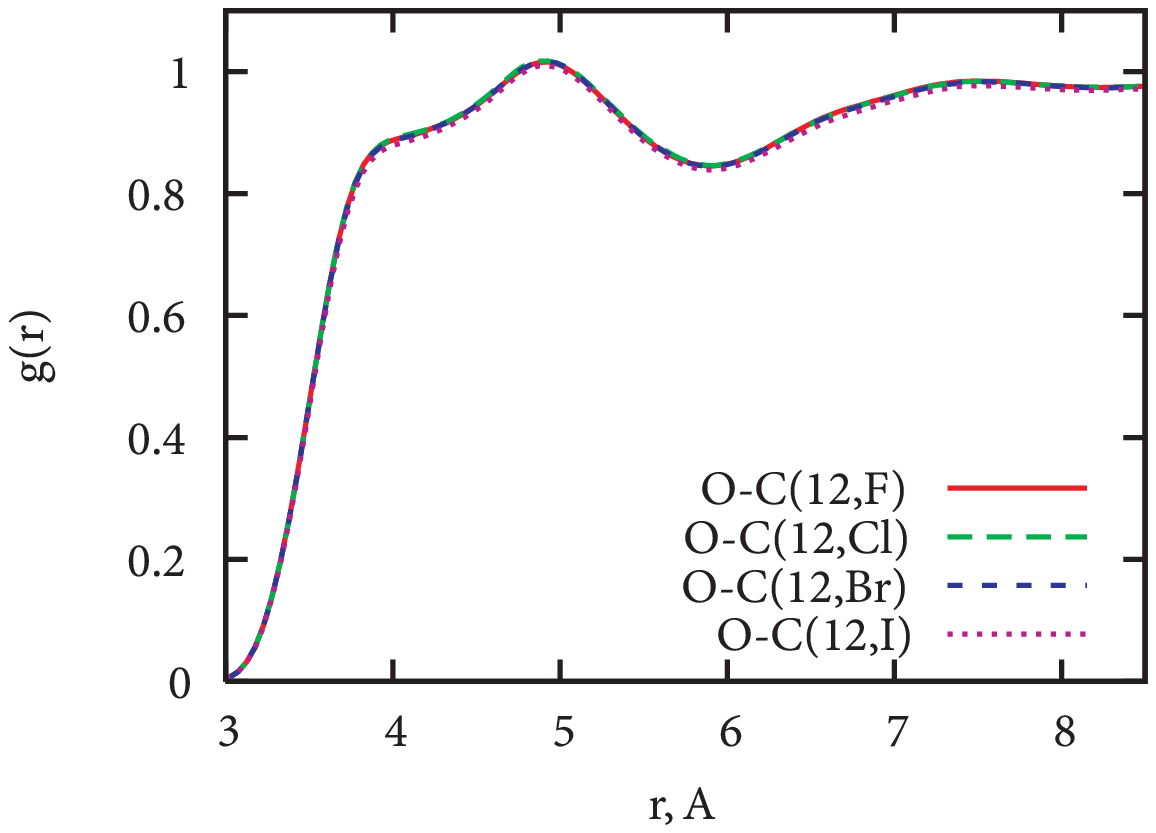}
}
\caption{(Color online) Oxygen-nitrogen (left-hand) and oxygen-carbon (right-hand) RDFs for molecule
with twelve methylene groups, C$_{12}$. The results for
fluorides are denoted by full red lines, for chlorides by dashed
green lines, bromides are denoted by short-dashed blue, and
iodides by dotted magenta lines.}
\label{fig:o-c12}
\end{figure}

To document the effect of the counterion species on the
hydration of the surfactant we collected the oxygen-nitrogen
and oxygen-carbon (methylene group) radial distribution
functions. All the surfactans studied here exhibit the same
trends as far as the counterion species is concerned, so only
the data for the longest one, C$_{12}$, are presented in more
detail (see figure~\ref{fig:o-c12}).
Oxygen-nitrogen radial distribution functions: the position of
the first peak remains unchanged, but its height slightly
decreases along the sequence F/Cl/Br/I. This result indicates
that large and loosely hydrated counterions win against waters
in competition for the vacancies near surfactant ions. A
similar effect has been earlier observed in the studies of 3,3-
and 6,6-ionene oligomers in
water~\cite{Druchok2008,Druchok2009}.

The right hand panel of figure~\ref{fig:o-c12} presents the
oxygen-carbon radial distribution functions. The differences
between the radial distribution functions belonging to
different counterions are even smaller here. It is the iodide
radial distribution function which has the smallest peak,
confirming the observation mentioned above. Finally, the shapes
of the oxygen-nitrogen and oxygen-carbon (methylene group)
radial distribution functions are the same as observed before
for $x,y$-ionene oligomers in
water~\cite{Druchok2008,Druchok2009}. Much weaker hydration
(drying) of the methylene groups in comparison with the
positively charged nitrogen group is clearly visible.

We expect for the positively charged surfactant head to orient
neighboring water molecules in such a way that an oxygen points
toward, and hydrogens away from the head. In addition, we are
interested in the orientation of water molecules near the hydrophobic
tail. A water molecule is considered to be a part of the
surfactant hydration shell if the distance between oxygen and
nitrogen is less than 6.3~{\AA}, or the distance between the oxygen
and tail carbon is less than 4.3~{\AA}. Notice that the carbons
attached directly to nitrogen are not included in the
statistics. These distances are the positions of the
corresponding first minima of the nitrogen-oxygen and
carbon-oxygen RDFs (figure~\ref{fig:o-c12}). We monitored the
angle between the dipole moment of a (hydrating) water molecule
and a vector pointing from oxygen to the nearest carbon or
nitrogen. Following this definition, we accumulated two types of
angular distribution functions. The first one describes orientations
of hydrating waters in the vicinity of nitrogen (charged head), and
the second one near the carbons of hydrophobic tail. To
illustrate the difference, we discuss these data in terms of
average angles. It is interesting that the angles averaged over
the waters near the ``nitrogen'' group show little dependence on
the counterion type, or on the length of the hydrophobic tail. The
variation of this angle is in the range between 97 to
100$^\circ$. Contrary to this, the orientation distribution of
the waters near ``carbons'' demonstrates a spread from 91$^\circ$
for C$_{12}$ to 97$^\circ$ for C$_3$. The result reflects the
fact that for shorter surfactants with higher charge density,
hydrating waters are more affected than for the longer
(C$_{12}$) ones.

\section{Discussion}

The quaternary ammonium group is a part of many important
molecules. One such example are aliphatic $x,y$-ionenes, where
$x$ and $y$ denote the numbers of methylene groups between the
two adjacent quaternary nitrogens. These cationic
polyelectrolytes can be synthesized with different charge
density, varying $x,y$ numbers from 3 to 12 (see, for
example~\cite{sint1,sint2,sintcha1,sintcha2}). Recently in a
series of papers we had studied aqueous solutions of
$x,y$-ionenes using different experimental
methods~\cite{Cebasek2011,Boncina2012,Serucnik2012}. In
addition, the molecular dynamics simulations of the 3,3- and
6,6-ionenes were performed~\cite{Druchok2008,Druchok2009}.

The repeating monomer unit of the 3,3-ionene polyelectrolyte
is composed of the quaternary ammonium group followed by three
methylene groups. Similarly, the repeating unit of
12,12-ionene is the quaternary ammonium group followed by
twelve CH$_2$ groups. It is quite clear from this that our
model C$_{12}$ surfactant can be ``identified'' as a monomer
unit (with an exception of the terminal hydrogen) of the
12,12-ionene polycation. With the same logic, the model C$_3$
surfactant is a building block of the 3,3-ionene, while the
C$_5$ and C$_8$ roughly correspond to the 6,6-ionene. This
makes it worthwhile to compare the present simulations with
those for $3,3$- and $6,6$-ionene oligoions in explicit
water. Unfortunately, no molecular dynamic simulations for the
$12,12$-ionene solutions exist so far.

Here, we compare the pair distribution functions shown in
figures~\ref{fig:n-x} and \ref{fig:c-x} with those published in
references~\cite{Druchok2008} (see figures~4 and 5 of that
paper), and~\cite{Druchok2009} (figure~3). Considering that the
force field used in both calculations is very similar, more or
less similar results are expected. The eventually observed
differences between the two results could only be attributed to
an accumulation of the charge on the oligoion with six monomer
units. The heights of the peaks in figure~4 (see
reference~\cite{Druchok2008}) follow the trend
$\textrm{NaF}<\textrm{NaCl}<\textrm{NaBr}<\textrm{NaI}$. The same holds true
for the carbon-counterion RDFs shown in figure~5 of the same
paper, except that now the iodide and bromide peaks are very
close to each other. In other words, the trends exhibited by
``monomers'' are also reproduced by short oligoions in water.
Useful information on the specific ion effects in
polyelectrolyte solutions may often be obtained by studying
monomer systems, as it was demonstrated for salts of
para-toluene(sulphonic) acid~\cite{Otrin}, and for
tetraalkylammonium halides in water (see reference~\cite{Krienke2009}, and the references therein). A word of
caution is needed with respect to this statement. Recent
measurements by \v Ceba\v sek and coworkers~\cite{Cebasek2013}
show that for sufficiently hydrophobic 12,12-ionenes, the
trends with respect to the nature of counterion (Hofmeister
series) may be reversed in comparison with the more charged
(less hydrophobic) 3,3-ionenes.

The experimental results for the alkyltrimethylammonium
surfactant solutions, which partially include the pre-micellar
region, have been presented by several
authors~\cite{Birch,Delisi1988,Jakubowska2008,Jakubowska2010}.
Since the critical micelle concentrations for surfactants
with long chains (C$_{12}$ and more) are very low, the data are
not collected systematically. Among these studies it is worth
mentioning the work by Jakubowska~\cite{Jakubowska2010} who
used mass spectrometry to investigate the affinity of
counterions to surfactant monomers in the gas phase. She
examined the sodium hexadecyl-N,N,N-trimethyl ammonium bromide
in the presence of various salts. The results suggested the
following ordering in the affinity of counterions for this
surfactant: $\rm{F}^-<\rm{Cl}^-<\rm{NO}_3^-$.

Enthalpies of dilution of various surfactants have been measured in
pre-micellar region by Birch and Hall~\cite{Birch};
see figure~1 of their paper. The authors compare their results
with the results by the Debye-H\"{u}ckel limiting law.
Surfactant ions (alkyltrimethyl ammonium bromides, C$_n$TAB)
with a number of carbon atoms $n$ from 6 to 12 were examined.
The surfactants exhibit both positive and negative deviations
from the limiting law. The ordering of the surfactants, from
negative to positive deviations is:
$\rm{C_6TAB}<\rm{C_8TAB}<\rm{C_{10}TAB}<\rm{C_{12}TAB}$. This
finding appears to be in qualitative agreement with the results
for counterion-carbon coordination number presented in
figure~\ref{fig:c-x_n}.

\section{Conclusions}

The explicit water molecular dynamics simulations of dilute
solutions of model alkyltrimethylammonium surfactant ions
(the number of carbon atoms in the tail is 3, 5, 8, 10, and 12) in
mixture with NaF, NaCl, NaBr, and NaI, are performed. The
results are presented in the form of relevant radial
distribution functions; in addition, the running coordination
numbers for the counterion-carbon distributions are evaluated.
The nitrogen-counterion correlations seem to primarily depend
on the Coulomb interaction. On the other hand, the
carbon-counterion distribution coordination numbers seem to be
also affected  by the van der Waals forces. The molecular
dynamics results presented here are consistent with the
experimental data for alkyltrimethyl ammonium salts taken in
the pre-micellar region.  Furthermore, the results agree with
similar simulations performed for aliphatic $x,y$-ionene
solutions, and with experimental findings for the surfactant
and polyelectrolyte systems containing a quaternary ammonium
group.

\section*{Acknowledgements}

This study was supported by the
Slovenian Research Agency fund (ARRS) through the Program
0103--0201, and Project J1--4148.


%
%

\ukrainianpart

\title{Взаємодія модельних іонів алкілтриметиламонію з іонами лужногалоїдних солей:
моделювання методом молекулярної динаміки із явно врахованими молекулами води}
\author{М.~Дручок\refaddr{label1},
        Ч. Подліпнік\refaddr{label2},
        В.~Влахи\refaddr{label2}}
\addresses{
\addr{label1} Інститут фізики конденсованих систем НАН України, вул. І.~Свєнціцького~1, 79011 Львів, Україна
\addr{label2} Факультет хімії та хімічної технології, Університет Любляни, вул. Ашкерчева 5, 1000 Любляна, Словенія
}

\makeukrtitle

\begin{abstract}
\tolerance=3000%
У роботі з допомогою методу молекулярної динаміки проведено моделювання
низькоконцентрованого розчину іонів алкілтриметиламонію (з кількістю
метиленових груп у ланцюгу 3, 5, 8, 10, 12) у суміші із солями NaF,
NaCl, NaBr або NaI при температурі 298~K. Для опису води використано модель SPC/E.
Результати представлено у формі низки
радіальних функцій розподілу між атомами азоту чи вуглецю (із груп CH$_2$)
алкілтриметиламонію та контріонами розчину. Для детальнішого висвітлення
результатів також наведено біжучі координаційні числа між атомами
вуглецю та контріонами. Виявлено, що контріони I$^-$ демонструють найвищу,
а F$^-$ найнижчу здатність асоціювати із іонами алкілтриметиламонію.
Огляд результатів проведено у світлі наявних експериментальних та
теоретичних даних для цих чи подібних систем.
\keywords сурфактанти, солі алкілтриметиламонію, лужногалоїдні солі, зв'язування іонів, молекулярна динаміка

\end{abstract}

\end{document}